# Enhancement and speed-up of carrier dynamics in a dielectric nanocavity with deep sub-wavelength confinement

Gaoneng Dong[1,2,*], Ali Nawaz Babar[1,2,*], Rasmus Ellebæk Christiansen[2,3], Søren Engelberth Hansen[1,2], Søren Stobbe[1,2]✉, Yi Yu[1,2]✉ & Jesper Mørk[1,2]✉

[1]*Department of Electrical and Photonics Engineering, Technical University of Denmark, 2800 Kgs. Lyngby, Denmark.*

[2]*NanoPhoton – Center for Nanophotonics, Technical University of Denmark, 2800 Kgs. Lyngby, Denmark.*

[3]*Department of Civil and Mechanical Engineering, Technical University of Denmark, Nils Koppels Allé, 2800 Kgs. Lyngby, Denmark.*

[*]*These authors contribute equally to this work.*

✉*E-mail: ssto@dtu.dk; yiyu@dtu.dk; jesm@dtu.dk*

**The emergence of dielectric bowtie cavities enable optical confinement with ultrahigh quality factor and ultra-small optical mode volumes with perspectives for enhanced light-matter interaction. Experimental work has so far emphasized the realization of these nanocavities. Here, we experimentally investigate the ultrafast dynamics of a topology-optimized dielectric (silicon) bowtie nanocavity, with device dimensions down to 12 nm, that localizes light to a mode volume deep below the so-called diffraction limit given by the half-wavelength cubed. This strong spatial light concentration is shown to significantly enhance the carrier generation rate through two-photon absorption, as well as reducing the time it takes for the carriers to recover. A diffusion time below 1 ps is achieved for the bowtie cavity, which is more than an order of magnitude smaller than for a conventional microcavity. Additionally, parametric effects due to coherent interactions between pump and probe signals are also enhanced in the bowtie cavity, leading to an improved extinction ratio. These results demonstrate important fundamental advantages of dielectric bowtie cavities compared to conventional point-defect cavities, laying a foundation for novel low-power and ultrafast optical devices, including switches and modulators.**

Optical microcavities[1] can significantly enhance the interaction between light and matter by providing strong spatial and spectral confinement, which is important for many applications, such as nanolasers[2-8], nonlinear photonics[9-14], sensing[15], cavity quantum electrodynamics[16-19], and optomechanics[20-22]. Traditional dielectric cavities, such as photonic crystal cavities, provide strong spectral confinement, while the spatial localization for such dielectric cavities for long was considered to be limited to mode volumes, $V_p$, larger than the so-called diffraction limit, $V_\lambda = (\lambda/(2n))^3$, with $\lambda$ and $n$ being the vacuum wavelength and material refractive index, respectively. Plasmonic cavities[23-25], on the other hand, are well-known to confine light deep below the wavelength, albeit at the cost of ohmic losses, which limit the spectral confinement[26]. Recently, however, a new class of nanocavities has emerged, so-called bowtie cavities[27-32], with experimentally demonstrated mode volumes down to $V_p = 0.08 V_\lambda$[33]. While the main argument for realizing such extreme dielectric confinement (EDC) cavities is to enhance the interaction between light and matter, experimental investigations of such effects are missing.

In this work, we experimentally study the dynamics of a bowtie cavity with deep sub-wavelength confinement of light. The cavity is designed using fabrication-constrained topology optimization while maintaining efficient waveguide-cavity coupling. The fabricated devices are characterized using a heterodyne pump-probe technique and show that, compared to a reference cavity with conventional confinement, the bowtie cavity not only enhances the strength of optical switching but also reduces the recovery time. We analyze the bowtie cavities using temporal coupled-mode theory and simulations of spatio-temporal carrier dynamics. The comparison with the experimental data shows that both incoherent processes due to enhanced two-photon absorption (TPA) as well as parametric interactions due to four-wave mixing between pump and probe are enhanced in the bowtie cavity. A diffusion time below 1 ps is achieved for the bowtie cavity, which is more than an order of magnitude smaller than for a conventional microcavity. Since the light-matter interaction is enhanced using a small optical mode volume rather than a high quality factor, the bandwidth of the device is not compromised, thus facilitating the development of high-speed applications, such as all-optical switching[11,13] and modulators[34-36].

**Structure and switching dynamics**

The bowtie cavity, designed using topology optimization[37] employing appropriate fabrication constraints[38] (see Methods and Supplement S1), is realized in a nanobeam geometry. It features an electromagnetic field profile with a strongly localized hotspot in the region between the two tips of the bowtie, as illustrated in Figure 1a (i). The nanobeam bowtie cavity distinguishes itself from previous two-dimensional (2D) structures[33,39] with two major benefits. Firstly, it seamlessly integrates input and output waveguides, which can be co-optimized with the cavity. In contrast to other nanobeam implementations[40], the design features a single rather than multiple hotspots, making it better suited for switching applications. The waveguide integration results in a transmission rate of 91% (−0.41 dB) at the resonant frequency. Secondly, the nanobeam has a footprint of just 3.0 µm², which is 20-30% of that occupied by previous 2D bowtie cavities[33,39].

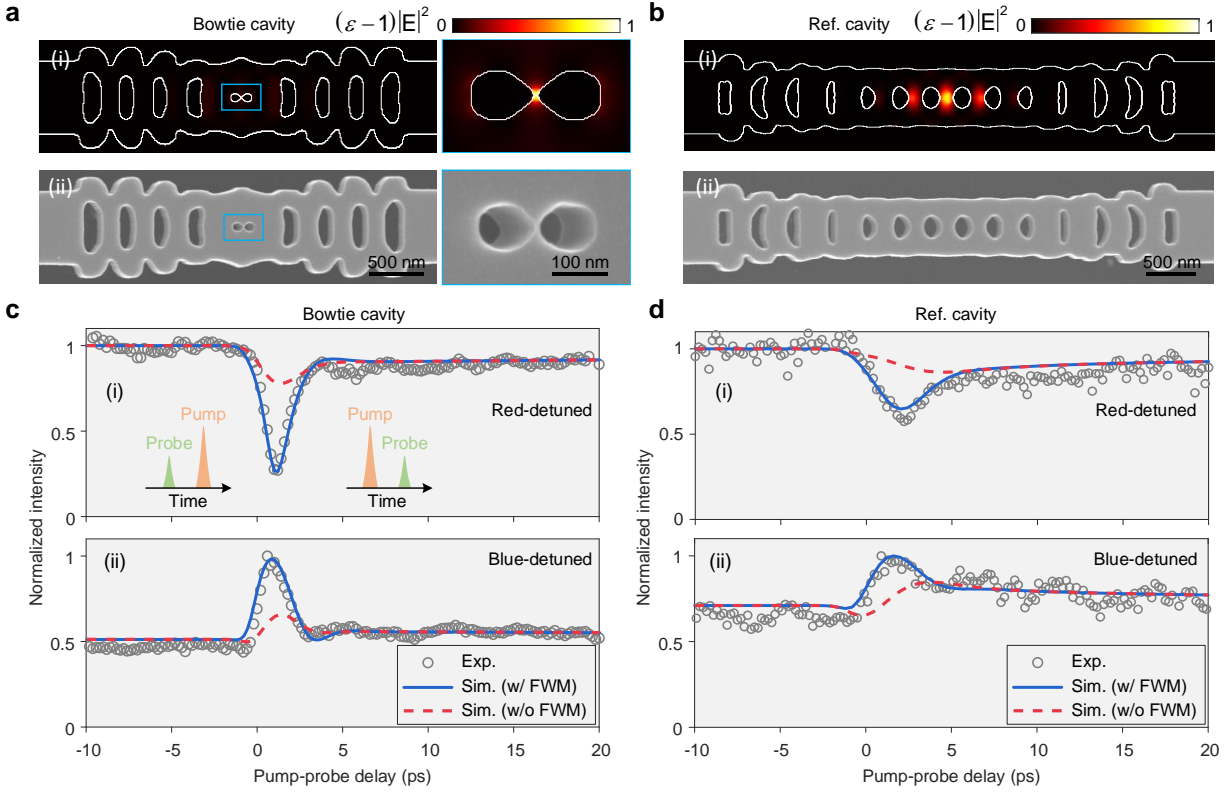

**Fig. 1 | Ultrafast dynamics of a nanobeam bowtie cavity**. **a**, (i) Normalized electric field intensity of the mode of the nanobeam bowtie cavity in the central plane. The inset on the right is an enlarged view of the central bowtie region. (ii) SEM images of the nanobeam bowtie cavity. **b**, (i) Normalized electric field intensity of the mode of the reference (Ref.) cavity in the central plane. (ii) SEM image of the reference cavity. **c**, **d**, Representative (normalized) probe transmission dynamics as a function of the pump-probe delay for a red-detuned (i) and blue-detuned (ii) probe wavelength in the nanobeam bowtie cavity (**c**) and reference cavity (**d**). In (**c**) and (**d**), the gray circle marks are the experimental (Exp.) results, while the solid blue and dashed red curves represent the simulations (Sim.) with (w/) and without (w/o) four-wave mixing (FWM), respectively. The pump pulse energy coupled into the waveguide is 255 fJ in (**c**) and (**d**).

In order to enable a one-to-one comparison of the dynamics of the bowtie cavity to conventional point-defect cavities, we also designed a reference nanobeam cavity in the same material platform (Fig. 1b (i)). The reference cavity has characteristics comparable to conventional nanobeam cavities[41], including the electric field distribution, quality factor, and mode volumes, while maintaining a similar footprint. These similarities extend to the carrier dynamics, which are on par with those in the conventional nanobeam cavities (see Methods and Supplement S2). It is emphasized that the fabrication uncertainties were incorporated into the topology optimization for both cavities, resulting in fabricated geometries (Figs. 1a (ii) and 1b (ii)) that closely mirror the designed structures (Figs. 1a (i) and 1b (i)).

The temporal dynamics of the devices were measured using a short-pulse pump-probe technique that allows co-polarized pump and probe pulses by using heterodyne detection[42] (see Methods and Supplement S4). The measured total quality factor and peak transmission of the characterized bowtie (reference) cavity are 700 (1200) and −0.82 dB (−0.47 dB), respectively, corresponding to an intrinsic quality factor of 7800 (23000) (see Supplement S5). Representative dynamic measurements are shown in Figs. 1c and 1d for the bowtie cavity and the reference cavity, respectively, considering both cases of red and blue detuning of the probe with respect to the pump (aligned to the cold cavity resonance). For the bowtie cavity, we observe a fast decrease (increase) in the transmittance of the red-detuned (blue-detuned) probe light after the arrival of an intense pump pulse with a pulse width of 0.93 ps. After the pump pulse, the probe transmittance recovers almost entirely within 2 ps. The amplitude of the slowly varying tail amounts to only one-tenth of the total change, and the extinction ratio is −5.7 dB for the red-detuned probe wavelength. In contrast, the pump-probe signal for the reference cavity shows an extinction ratio of −2.3 dB at the same injected pump energy, and has a significantly longer tail (Fig. 1d).

In order to interpret the data, we model the probe transmission dynamics using coupled-mode theory, accounting for the generation of free carriers in the cavity region by TPA, as well as coherent wave mixing between the pump and probe pulses due to oscillations of the carrier density mediated by beating[43,44] (see Methods and Supplement S3). The simulated results, solid blue curves in Figs. 1c and 1d, agree well with measurements when using the parameter values given in Table S3 of the supplement. If parametric processes are neglected in the model, represented by the dashed red curves, the experimental data cannot be fitted by the theory in a time interval around zero delay, where the pump and probe pulses temporally overlap. On the other hand, the slower components are still well represented by the model.

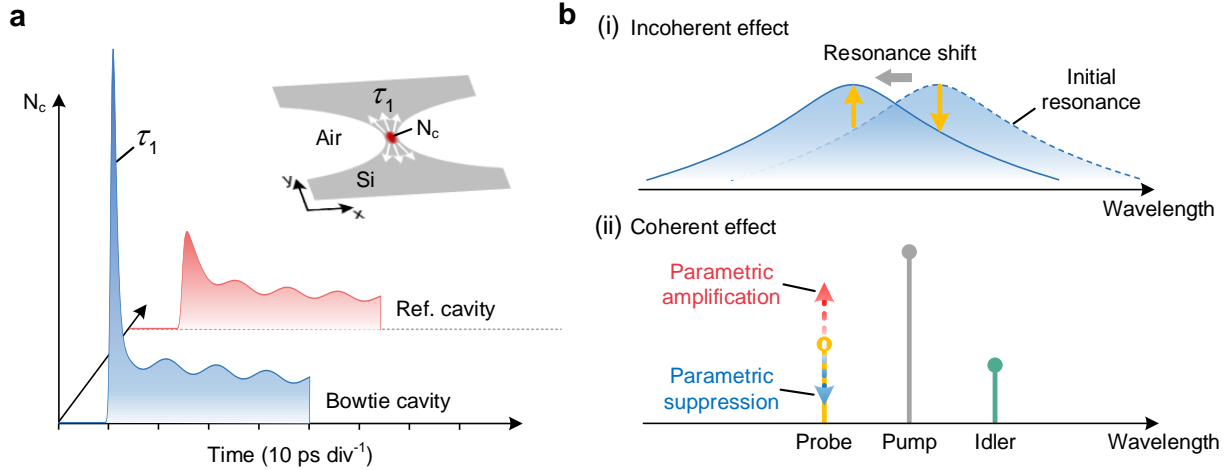

**Fig. 2 | Physical mechanisms. a**, Time evolution (calculated with the parameters given in Table S3 of the supplement, but using the same quality factor ($Q_t = 700$) and pump pulse width (0.93 ps) for both cavities) of the mode-averaged carrier density, $N_c$, in the bowtie cavity (blue shaded area) and reference cavity (red shaded area). The inset illustrates the carrier diffusion process in the central plane of the cavity mode region of the bowtie cavity. **b**, Working principle in the frequency domain. (i) Probe transmission change caused by the frequency shift of the cavity resonance. (ii) Probe transmission change induced by the parametric amplification or suppression process.

Based on these results, a simple physical picture emerges for the dynamics induced by the pump, cf. Figure 2. The main effect is the generation of free carriers by TPA of the pump pulse. Due to free-carrier dispersion and band-filling, these carriers change the refractive index in the region where they are generated, i.e., in the nanocavity. This leads to an incoherent and a coherent (parametric) change of the probe transmission, cf. Fig. 2(b).

The incoherent effect is the shift of the resonance frequency of the cavity, cf. Fig. 2b (i), which changes the transmission of the probe field. Since the cavity always blue shifts with increased carrier density, the sign of the probe amplitude change depends on the position of the probe wavelength relative to the cavity resonance. The relaxation time of the probe amplitude change is instead determined by the temporal variation of the spatial overlap between the carrier distribution and the probe mode profile.

The parametric effect is caused by the temporal oscillation of the free carrier density due to beating between the co-polarized pump and probe fields, cf. Fig. 2a. This establishes a temporal grating that scatters photons between the pump and probe wavelengths, accompanied by the generation of an idler signal, cf. Fig. 2b (ii). Whether the probe field is amplified or attenuated depends on the phase-matching condition between the pump light, the probe light, and the oscillating carrier density[43]. Since the pump and probe fields only beat when the pulses overlap, this effect is limited to a time window around zero pump-probe delay.

The resulting probe dynamics is the combination of these effects. Both the larger amplitude variation and the shorter time scale of the dynamics seen in the bowtie cavity compared to the reference cavity originate from the stronger spatial localization of light in the bowtie region. The localization enhances the carrier generation rate since this happens via a nonlinear process (TPA in our case), and the time scale for re-equilibration after the pump pulse shortens due to the larger spatial gradient of the carrier distribution, which enhances the diffusion process. Next, we quantify these effects through an analysis of the characteristic mode volumes.

## Mode volumes and carrier diffusion

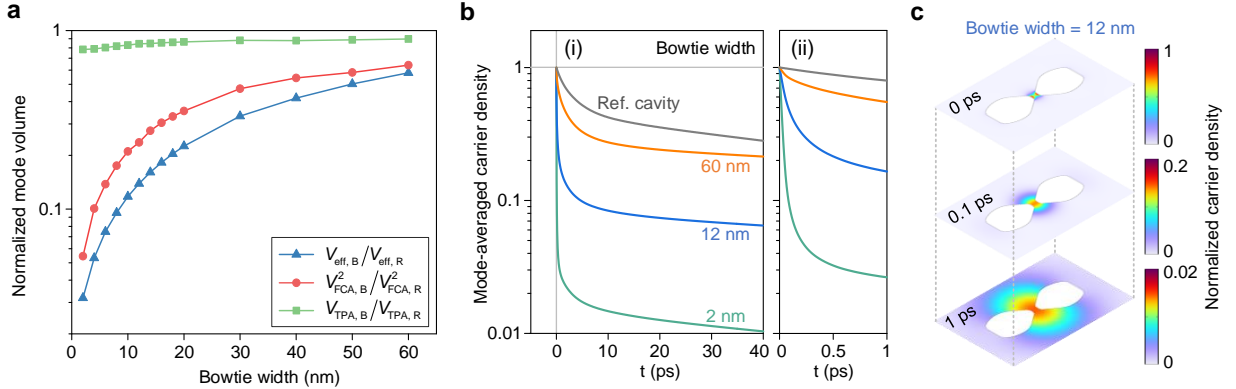

**Fig. 3 | Mode volumes and carrier diffusion**. **a**, Calculated effective mode volume, $V_{eff}$, square of free-carrier absorption (FCA) mode volume, $V_{FCA}$, and two-photon absorption (TPA) mode volume, $V_{TPA}$, of the nanobeam bowtie cavity ($B$) versus width of the bowtie bridge. These mode volumes are normalized by corresponding mode volumes of the reference cavity ($R$). **b**, Calculated (normalized) carrier relaxation process for the nanobeam bowtie cavity with different bowtie widths (2 nm, 12 nm, and 60 nm) and the reference cavity. (ii) shows a zoom-in on the short time-scale dynamics in (i). **c**, Calculated free carrier distribution in the central plane of the bowtie region (bowtie width = 12 nm) at different times (0 ps, 0.1 ps, and 1 ps). The carrier distributions are normalized by the maximum carrier density at 0 ps.

The strengths of linear and nonlinear optical processes in a cavity scale with different mode volumes[45,46]. To quantify these, we calculate the effective (linear) mode volume, $V_{eff}$, as well as the nonlinear mode volumes, i.e., the free-carrier absorption (FCA) mode volume, $V_{FCA}$, and the TPA mode volume, $V_{TPA}$[45,46,33] (see Supplement S2), as shown in Figure 3a.

It is important to recognize that the rate, $G_N$, of carrier generation due to TPA scales inversely with the square of the FCA mode volume, i.e., $G_N \propto 1/V_{FCA}^2$ [13]. Conversely, the TPA loss, $\gamma_{TPA}$, is inversely proportional to the TPA mode volume, i.e., $\gamma_{TPA} \propto 1/V_{TPA}$ (see Supplement S3). Across a wide range (2 – 60 nm) of bowtie widths, the FCA mode volume is significantly smaller than that of the reference cavity. Notably, as the width of the bowtie bridge narrows, there is a pronounced reduction in the FCA mode volume. Consequently, when subjected to identical pump pulse energies, the carrier generation rate is significantly higher in the bowtie cavity than the reference cavity, as illustrated in Fig. 2a. This leads to a larger resonance shift and a larger amplitude of the parametric processes (Fig. 1c). Remarkably, while the FCA mode volume of the bowtie cavity diminishes rapidly with the bowtie width, the TPA mode volume exhibits only a minor reduction and is on par with that of the reference cavity. This suggests that the bowtie cavity imparts a larger dynamical change of the transmission without a corresponding increase in TPA loss. Although the FCA loss in the bowtie cavity is also increased due to the smaller $V_{FCA}$, this only results in a small decrease in the extinction ratio according to our calculations (see Supplement S7).

The effective mode volume does not directly impact the amplitude of the dynamical change of the probe, but is important for the speed of the carrier diffusion process, as previously demonstrated in photonic crystal nanocavities[11,13,47]. The cavity with a smaller effective mode volume results in a faster carrier relaxation process due to the larger spatial gradients of free carriers in the cavity mode region.

Figure 3b shows the calculated carrier relaxation processes for the bowtie cavity with different bowtie widths and the reference cavity using the ambipolar-diffusion model[47,48]. The results show that the carriers in the bowtie cavity diffuse out of the cavity mode region much faster than the reference cavity over a wide range of bowtie widths. This advantage of the bowtie cavity becomes more significant as the bowtie width decreases. Representative time-dependent carrier distributions of the bowtie cavity (bowtie width = 12 nm) show that the maximum carrier density decreases by more than 50 times within 1 ps (Fig. 3c).

The carrier relaxation process consists of a fast and a slow component (Fig. 3b). According to our calculations (see Supplement S2), both the fast and slow components are mainly determined by carrier diffusion when the surface recombination rate is lower than $10^5$ cm/s. For the experimentally characterized bowtie cavity with a bowtie width of 12 nm, the calculated fast diffusion time is approximately 0.3 ps, while this value drops to tens of fs when the bowtie width is reduced to 2 nm. In contrast, the fast diffusion time of the reference cavity is 6 ps, which is more than one order of magnitude larger than that of the bowtie cavity. It is noteworthy that the bowtie cavity not only has a shorter fast diffusion time than the reference cavity, but also has a larger relative amplitude of the fast diffusion component. This is why the slow recovery component, reflecting slow diffusion and non-radiative recombination, is strongly suppressed in the bowtie cavity compared to the reference cavity (Figs. 1c and 1d).

## Wavelength-dependent dynamics

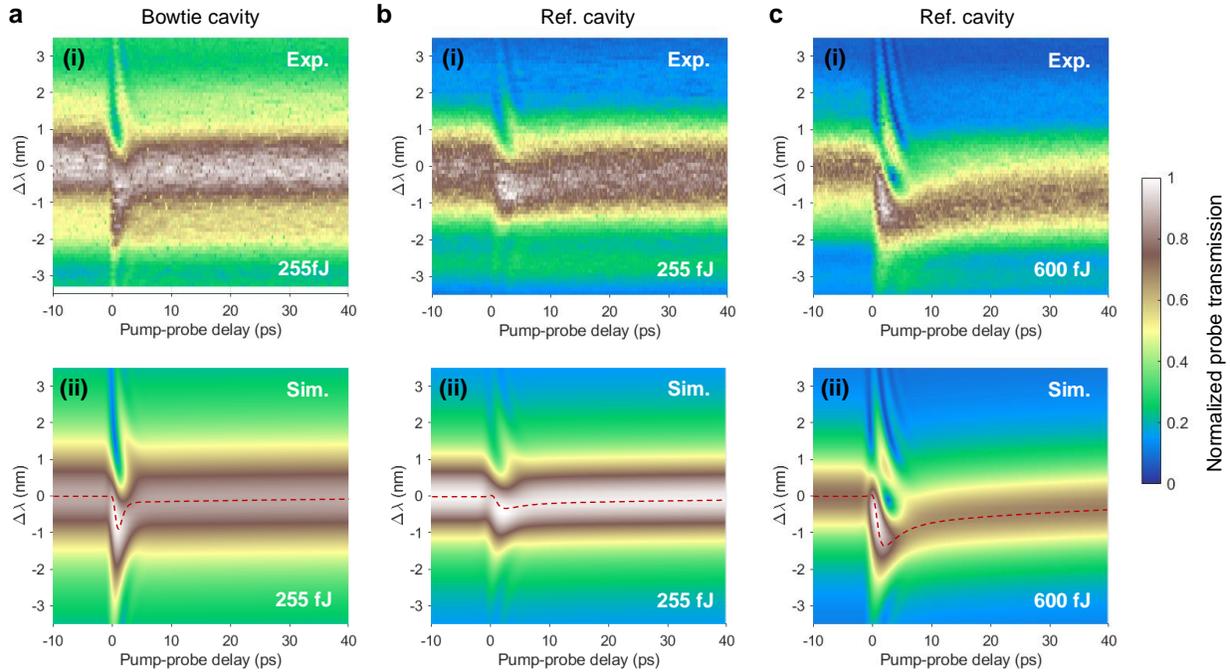

**Fig. 4 | Wavelength-dependent dynamics**. **a**, Measured (i) and calculated (ii) probe transmission dynamics of the nanobeam bowtie cavity for a pump pulse energy of 255 fJ. **b**, Measured (i) and calculated (ii) probe transmission dynamics of the reference cavity for a pump pulse energy of 255 fJ. **c**, Measured (i) and calculated (ii) probe transmission dynamics of the reference cavity for a pump pulse energy of 600 fJ. The red dashed curves in **a** (ii), **b** (ii), and **c** (ii) represent the corresponding cavity resonance shifts.

To further investigate the dynamics, in particular the differences between the bowtie and the reference cavity, we compared experimental and theoretical results for the wavelength and time-dependent variation of the probe signal, cf. Figure 4. Figure 4a shows results for the bowtie cavity for a fixed pump wavelength coinciding with the cold cavity resonance and an injected pump pulse energy of 255 fJ at the input waveguide. Good agreement between experiment and theory is obtained using the set of parameters given in Table S3 of the supplement. The red dashed curve represents the resonance shift of the bowtie cavity, indicating a switching window of 1.5 ps as evaluated from the full width at half the maximum resonance shift. The fast diffusion time used for the fitting is 0.6 ps, which is comparable to the numerically simulated result of 0.3 ps obtained using the ambipolar-diffusion model. The simulated probe transmission dynamics in the absence of the parametric process are also shown (see Supplement S3), indicating that the ultrafast probe transmission increase (decrease) is dominated by the parametric process while the slow recovery tail depends on the cavity resonance shift. The significance of the parametric process in a time-window around zero pump-probe delay was confirmed experimentally by using the heterodyne technique to measure the idler signal[43]. As shown in Supplement S6, the idler signal only exists in the time-window around zero pump-probe delay, confirming the coherent nature of this process.

Results for the reference cavity (Fig. 4b), also conducted with the pump pulse in resonance with the cavity and for the same pump pulse energy of 255 fJ, show that the probe change due to both the parametric process and the cavity resonance shift, represented by the red dashed curves, are much weaker than for the bowtie cavity, despite the quality factor of the cavity being significantly higher. Although the resonance shift is relatively small for the reference cavity, the slow recovery process is clearly visible (Fig. 4b). The slow-down of the diffusion process is even more apparent if the experiment is carried out for higher pump pulse energy (Fig. 4c). The features seen in Fig. 4b and 4c for the reference cavity are very similar to those observed in a photonic crystal H0 cavity[44], but differ significantly from the bowtie cavity. The estimated switching window for the bowtie cavity is approximately an order of magnitude smaller than the 18-ps window observed in the reference cavity.

**Discussion**

Although the pump power required to achieve a given resonance shift is already lower for the bowtie cavity with a bridge width of 12 nm than for the reference cavity, the pump power can be further reduced by reducing the bowtie width due to the scaling with the nonlinear mode volume, $V_{\text{FCA}}$. It should be noted, though, that when the bowtie width decreases, the faster diffusion of free carriers counteracts the accumulation of carriers in the modal region, thereby limiting the reduction of the energy consumption. However, on the positive side, the fast diffusion reduces the amplitude of the slow tail. A bowtie cavity with a smaller bowtie width may be achieved through thickness-controlled surface oxidation and subsequent oxide layer removal[49,50]. The pump power can also be significantly reduced by using other materials, such as InP[13] and GaAs[10], whose TPA coefficients are more than one order of magnitude larger than Si.

In summary, we experimentally investigate the dynamics of a dielectric cavity featuring a bowtie nanostructure that enables sub-wavelength confinement of light inside the semiconductor material. Both theoretical and experimental results show that the bowtie cavity significantly improves the all-optical switching dynamics, owing to an ultra-small linear mode volume as well as an ultra-small nonlinear mode volume. The smaller linear mode volume speeds up the carrier diffusion process and reduces the slow recovery tail commonly associated with switches. The smaller nonlinear mode volume improves the switching contrast through both coherent and incoherent effects. These results establish the dielectric bowtie cavity as a new and strong candidate for low-power and ultrafast optical devices, including switches[11,13,14] and modulators[34-36].

## Methods

**Device design**

The two waveguide-coupled cavities studied in this work are designed using topology optimization, including constraints dictated by the fabrication technology[38]. The bowtie cavity was designed to minimize the effective mode volume, $V_{eff}$. In brief, this was done using a classical time-harmonic model of the electromagnetic field by exciting the nanobeam with the lowest order propagating waveguide mode and tailoring the nanobeam geometry to maximize the resulting electric field amplitude at a point in the center of the beam, under the design constraint that the transmittance of the waveguide mode through the nanobeam remains above 90%. The design problems were solved using COMSOL Multiphysics executed on the DTU Computing Center HPC cluster[51]. More details are provided in Supplement S1.

**Device fabrication**

Our devices utilize a 220-nm thick crystalline (100) silicon device layer. The fabrication involves metal hard mask deposition, resist spin-coating, electron-beam lithography, reactive-ion etching, and selective vapor-phase hydrofluoric acid etching. The hard mask stack comprises a 50-nm CSAR resist, a 10-nm poly-silicon, and a 10-nm poly-chromium layer for etching high-aspect ratio features into the silicon device layer with high fidelity. The nanofabrication process starts with depositing a 10-nm chromium layer and a 10-nm poly-silicon layer using sputtering, followed by spin coating of 50-nm thin electron-sensitive resist CSAR. Then, electron-beam lithography is performed with a 100 keV JEOL JBX 9500FSZ tool. A current of I = 0.2 nA and a shot-pitch of 1 nm are used to expose nanocavities in the resist. Etching is performed by deep reactive-ion etching using the double-hard mask stack. The structures are suspended using an anhydrous vapor-phase hydrofluoric acid etch. More details are provided in Ref. 31.

**Pump-probe measurements**

A Ti:Sapphire pulsed laser (Spectra-Physics Maitai) combined with an optical parametric oscillator (Spectra-Physics Inspire HP) produced ~170-fs pulses at 1.54 μm with a 79.9 MHz repetition rate. Two acousto-optical modulators (AOM) are used to generate the pump and probe beams with frequency shifts of 85 MHz and 80 MHz, respectively, compared to the undefelected reference beam. A fiber-coupled waveshaper (Finisar Waveshaper 4000S/X) employing Gaussian filters is used to reshape the pump and probe spectra and compensate for dispersion, keeping the pulses transform-limited. The pump pulse width was chosen according to the cavity photon lifetime of each cavity to obtain the maximum extinction ratio for the switching dynamics (see Supplement S8). Two adjustable delay lines allowed timing adjustments between the pump, probe, and reference pulses. The pump and probe pulses, with transverse polarization, were coupled into the waveguide through a grating coupler with a –6 dB loss per facet. The transmitted mixed signal was detected using heterodyne cross-correlation with a 50:50 beam splitter. A lock-in amplifier (Stanford Research 844) extracted the probe or idler signal, locked to the corresponding beating frequency. More details are provided in Supplement S4.

**Numerical simulations**

The probe transmission dynamics are calculated using temporal coupled mode theory, which includes the effects of coherent beating between the pump and probe waves. The probe wave is considered as a small perturbation to the entire intracavity field. More details are provided in Supplement S3.

**Data Availability Statement**

The data used in this work are available from the corresponding author upon reasonable request.

## Acknowledgments

The authors gratefully acknowledge financial support from the Danish National Research Foundation (Grant no. DNRF147 NanoPhoton), the European Research Council (ERC) under the European Union Horizon 2020 Research and Innovation Programme (Grants no. 834410 FANO and no. 101045396 SPOTLIGHT)), Villum Fonden via the Young Investigator Program (Grant no. 42026 EXTREME), and Innovation Fund Denmark (Grant no. 0175-00022—NEXUS).

## Author Contributions

J.M., S.S., and Y.Y. initiated and supervised the project. G.D., supervised by Y.Y. and J.M., performed pump-probe measurements, theoretical simulations, and simulated mode volumes and carrier dynamics. A.N.B., supervised by S.S., fabricated the sample, performed SEM characterization, and contributed to the simulations of mode volumes and carrier dynamics. R.E.C. developed and carried out the topology optimization. S.E.H. designed the grating coupler. G.D., J.M., and Y.Y. wrote the manuscript with contributions and input from all authors.

## Competing interests

The authors declare no competing interests.

## Additional information

**Supplementary information**

## References

1. Vahala, K. J., "Optical microcavities," Nature **424**, 839-846 (2003).

2. Park, H.-G., et al., "Electrically driven single-cell photonic crystal laser," Science **305**, 1444 (2004).

3. Ellis, B., et al., "Ultralow-threshold electrically pumped quantum-dot photonic-crystal nanocavity laser," Nat. Photonics **5**, 297-300 (2011).

4. Jeong, K.-Y., et al., "Electrically driven nanobeam laser," Nat. Commun. **4**, 2822 (2013).

5. Takeda, K., et al., "Few-fJ/bit data transmissions using directly modulated lambda-scale embedded active region photonic-crystal lasers," Nat. Photonics **7**, 569-575 (2013).

6. Hill, M. T. & Gather, M. C., "Advances in small lasers," Nat. Photonics **8**, 908-918 (2014).

7. Crosnier, G., et al., "Hybrid indium phosphide-on-silicon nanolaser diode," Nat. Photonics **11**, 297-300 (2017).

8. Dimopoulos, E., et al., "Experimental demonstration of a nanolaser with a sub-µA threshold current," Optica **10**, 973 (2023).

9. Tanabe, T., et al., "All-optical switches on a silicon chip realized using photonic crystal nanocavities," Appl. Phys. Lett. **87**, 151112 (2005).

10. Husko, C., et al., "Ultrafast all-optical modulation in GaAs photonic crystal cavities," Appl. Phys. Lett. **94**, 021111 (2009).

11. Nozaki, K., et al., "Sub-femtojoule all-optical switching using a photonic-crystal nanocavity," Nat. Photonics **4**, 477-483 (2010).

12. Nozaki, K., et al., "Ultralow-power all-optical RAM based on nanocavities," Nat. Photonics **6**, 248-252 (2012).

13. Yu, Y., et al., "Switching characteristics of an InP photonic crystal nanocavity: Experiment and theory," Opt. Express **21**, 31047-31061 (2013).

14. Moille, G., et al., "Integrated all-optical switch with 10 ps time resolution enabled by ALD," Laser Photonics Rev. **10**, 409-419 (2016).

15. Shambat, G., et al., "Single-cell photonic nanocavity probes," Nano Lett. **13**, 4999-5005 (2013).

16. Noda, S., Fujita, M. & Asano, T., "Spontaneous-emission control by photonic crystals and nanocavities," Nat. Photonics **1**, 449-458 (2007).

17. Tiecke, T. G., et al., "Nanophotonic quantum phase switch with a single atom," Nature **508**, 241-244 (2014).

18. Lodahl, P., Mahmoodian, S. & Stobbe, S., "Interfacing single photons and single quantum dots with photonic nanostructures," Rev. Mod. Phys. **87**, 347-400 (2015).

19. Sipahigil, A., et al., "An integrated diamond nanophotonics platform for quantum-optical networks," Science **354**, 847-850 (2016).

20. Eichenfield, M., et al., "A picogram- and nanometre-scale photonic-crystal optomechanical cavity," Nature **459**, 550-555 (2009).

21. Safavi-Naeini, A. H., et al., "Electromagnetically induced transparency and slow light with optomechanics," Nature **472**, 69-73 (2011).

22. Cohen, J. D., et al., "Phonon counting and intensity interferometry of a nanomechanical resonator," Nature **520**, 522-525 (2015).

23. Oulton, R. F., et al., "Plasmon lasers at deep subwavelength scale," Nature **461**, 629-632 (2009).

24. Khajavikhan, M., et al., "Thresholdless nanoscale coaxial lasers," Nature **482**, 204-207 (2012).

25. Xu, J., et al., "Room-temperature low-threshold plasmonic nanolaser through mode-tailoring at communication wavelengths," Laser Photonics Rev., 2200322 (2022).

26. Hwang, M. S., et al., "Recent advances in nanocavities and their applications," Chem Commun (Camb) **57**, 4875-4885 (2021).

27. Gondarenko, A. & Lipson, M., "Low modal volume dipole-like dielectric slab resonator," Opt. Express **16**, 17689-17694 (2008).

28. Hu, S. & Weiss, S. M., "Design of photonic crystal cavities for extreme light concentration," ACS Photonics **3**, 1647-1653 (2016).

29. Choi, H., Heuck, M. & Englund, D., "Self-similar nanocavity design with ultrasmall mode volume for single-photon nonlinearities," Phys. Rev. Lett. **118**, 223605 (2017).

30. Wang, F., et al., "Maximizing the quality factor to mode volume ratio for ultra-small photonic crystal cavities," Appl. Phys. Lett. **113**, 241101 (2018).

31. Babar, A. N., et al., "Self-assembled photonic cavities with atomic-scale confinement," Nature **624**, 57-63 (2023).

32. Ouyang, Y. H., et al., "Singular dielectric nanolaser with atomic-scale field localization," Nature **632**, 287-293 (2024).

33. Albrechtsen, M., et al., "Nanometer-scale photon confinement in topology-optimized dielectric cavities," Nat. Commun. **13**, 6281 (2022).

34. Xu, Q., Schmidt, B., Pradhan, S. & Lipson, M., "Micrometre-scale silicon electro-optic modulator," Nature **435**, 325-327 (2005).

35. Li, M., et al., "Lithium niobate photonic-crystal electro-optic modulator," Nat. Commun. **11**, 4123 (2020).

36. Eppenberger, M., et al., "Resonant plasmonic micro-racetrack modulators with high bandwidth and high temperature tolerance," Nat. Photonics (2023).

37. Christiansen, R. E. & Sigmund, O., "Inverse design in photonics by topology optimization: tutorial," J. Opt. Soc. Am. B **38**, 496 (2021).

38. Christiansen, R. E., "Inverse design of optical mode converters by topology optimization: tutorial," J. Opt. **25**, 083501 (2023).

39. Xiong, M., et al., "Experimental realization of deep sub-wavelength confinement of light in a topology-optimized InP nanocavity," Opt. Mater. Express **14**, 397-406 (2024).

40. Hu, S., et al., "Experimental realization of deep-subwavelength confinement in dielectric optical resonators," Sci. Adv. **4**, eaat2355 (2018).

41. Deotare, P. B., et al., "High quality factor photonic crystal nanobeam cavities," Appl. Phys. Lett. **94**, 121106 (2009).

42. Mecozzi, A. & Mørk, J., "Theory of heterodyne pump–probe experiments with femtosecond pulses," J. Opt. Soc. Am. B **13**, 2437-2452 (1996).

43. Colman, P., Lunnemann, P., Yu, Y. & Mork, J., "Ultrafast coherent dynamics of a photonic crystal all-optical switch," Phys. Rev. Lett. **117**, 233901 (2016).

44. Lunnemann, P., Yu, Y., Joanesarson, K. & Mørk, J., "Ultrafast parametric process in a photonic-crystal nanocavity switch," Phys. Rev. A **99**, 053835 (2019).

45. Barclay, P. E., Srinivasan, K. & Painter, O., "Nonlinear response of silicon photonic crystal microresonators excited via an integrated waveguide and fiber taper," Opt. Express **13**, 801-820 (2005).

46. Johnson, T. J., Borselli, M. & Painter, O., "Self-induced optical modulation of the transmission through a high-Q silicon microdisk resonator," Opt. Express **14**, 817-831 (2006).

47. Moille, G., Combrié, S. & De Rossi, A., "Modeling of the carrier dynamics in nonlinear semiconductor nanoscale resonators," Phys. Rev. A **94**(2016).

48. Saldutti, M., et al., "Carrier diffusion in semiconductor nanoscale resonators," Phys. Rev. B **109**, 245301 (2024).

49. Lee, H. S., et al., "Local tuning of photonic crystal nanocavity modes by laser-assisted oxidation," Appl. Phys. Lett. **95**, 191109 (2009).

50. Chen, C. J., et al., "Selective tuning of high-Q silicon photonic crystal nanocavities via laser-assisted local oxidation," Opt. Express **19**, 12480-12489 (2011).

51. "DTU Computing Center, "DTU Computing Center resources," (2022).".

# Supplementary information for

# Enhancement and speed-up of carrier dynamics in a dielectric nanocavity with deep sub-wavelength confinement

Gaoneng Dong[1,2,*], Ali Nawaz Babar[1,2,*], Rasmus Ellebæk Christiansen[1,3], Søren Engelberth Hansen[1,2], Søren Stobbe[1,2]✉, Yi Yu[1,2]✉ & Jesper Mørk[1,2]✉

[1] *Department of Electrical and Photonics Engineering, Technical University of Denmark, 2800 Kgs. Lyngby, Denmark.*

[2] *NanoPhoton – Center for Nanophotonics, Technical University of Denmark, 2800 Kgs. Lyngby, Denmark.*

[3] *Department of Civil and Mechanical Engineering, Technical University of Denmark, Nils Koppels Allé, 2800 Kgs. Lyngby, Denmark.*

[*] *These authors contribute equally to this work.*

✉*E-mail: ssto@dtu.dk; yiyu@dtu.dk; jesm@dtu.dk*

**Contents**

## S1. Design of nanobeam bowtie cavity and reference cavity

The waveguide-coupled cavities studied in this work are designed using fabrication-constrained topology optimization [1] to operate resonantly at a particular wavelength. The physics model employed in the design process is derived from the classical Maxwell equations under the assumption of time-harmonic fields, leading to a single partial differential equation (s1) describing the electric field, **E**, in the modeling domain $\Omega_I$,

$$\nabla \times \nabla \times \mathbf{E}(\mathbf{r}) - \frac{\omega^2}{c^2}\varepsilon_r(\mathbf{r})\mathbf{E}(\mathbf{r}) = 0, \quad \mathbf{r} \in \Omega_I \subset \mathbb{R}^3, \tag{s1}$$

for a given material distribution determined by $\varepsilon(\mathbf{r}) = n(\mathbf{r})^2$, and free-space wavelength, $\lambda = \frac{2\pi c}{\omega}$. A design domain $\Omega_D$ is defined inside the model domain, in which the cavity under design resides. The design domain is connected to an input and an output waveguide. The model domain is bounded using first-order absorbing boundary conditions (s2). The model is driven by a port boundary condition exciting the fundamental waveguide mode propagating from the input waveguide (s3) through the design domain and into the output waveguide (s4). The field exiting the output waveguide is absorbed using another port boundary condition.

$$\mathbf{n} \times (\nabla \times \mathbf{E}(\mathbf{r})) - i\frac{\omega}{c}\varepsilon_r(\mathbf{r})\,\mathbf{n} \times (\mathbf{E}(\mathbf{r}) \times \mathbf{n}) = 0, \quad \mathbf{r} \in \Gamma_{\text{ABS}} \subset \mathbb{R}^3, \tag{s2}$$

$$\mathbf{n} \times (\nabla \times \mathbf{E}(\mathbf{r})) - ik_{\text{In}}\mathbf{n} \times (\mathbf{E}(\mathbf{r}) \times \mathbf{n}) = -2ik_{\text{in}}\mathbf{E}_{\text{in}}(\mathbf{r}), \quad \mathbf{r} \in \Gamma_{\text{In}} \subset \mathbb{R}^3, \tag{s3}$$

$$\mathbf{n} \times (\nabla \times \mathbf{E}(\mathbf{r})) - ik_{\text{Out}}\mathbf{n} \times (\mathbf{E}(\mathbf{r}) \times \mathbf{n}) = 0, \quad \mathbf{r} \in \Gamma_{\text{Out}} \subset \mathbb{R}^3. \tag{s4}$$

The model is reduced by exploiting the symmetries of the problem to only consider a fourth of the domain by imposing perfect electric conduction ($\mathbf{n} \times \mathbf{E} = \mathbf{0}$) and perfect magnetic conduction ($\mathbf{n} \times \mathbf{H} = \mathbf{0}$) as appropriate. The symmetry-reduced spatial configuration considered in the model is illustrated in Fig. S1.

The physics model is discretized using the finite element method employing first-order Nedelec elements [2] and solved to obtain the electric field for a given material distribution. The material distribution in the design domain, constituting the cavity under design, is modelled using voxelwise constant values, coinciding with the finite element mesh and controlled by a voxelwise constant design field. The in-plane design resolution is set at 5 nm x 5 nm away from $r_0$ (1 nm x 1 nm resolution near $r_0$) with five elements discretizing the device layer in the out-of-plane direction, $\xi(\mathbf{r}) = \sum_{j=1}^{N_e} \xi_j M_j(\mathbf{r})$, with $\xi_j$ being the design variable in the model. A planar design is assumed, thus the design variables are linked in the out-of-plane direction [3]. A standard filtering and thresholding approach with continuation (employing 200 design iterations per continuation step with the filter radius $r_f$, the threshold value $\eta$, and threshold sharpness $\beta$ given in Table S1) is used to regularize the design field [4]. The regularized field is used to linearly interpolate the relative permittivity in each voxel in the design domain between the background material (air) and the device material (silicon). In order to ensure that the design can be fabricated, a connectivity constraint [1] and a length-scale constraint [5] is enforced on the design geometry.

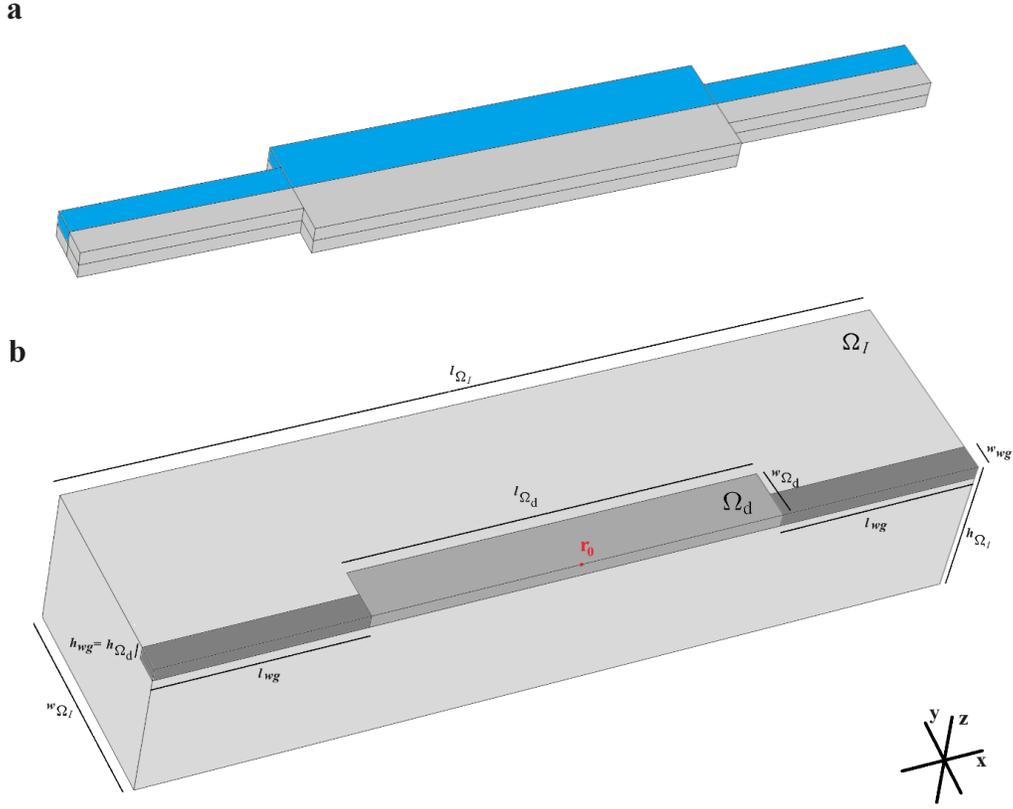

**Fig. S1**. **a,** Waveguide-coupled device with the mirror-symmetric subdomain highlighted in blue. **b,** Mirror symmetric modeling domain $\Omega_I$ with waveguide, design domain $\Omega_D$, and surrounding background (air), highlighting the target point $r_0$.

As the waveguide-coupled cavities are intended to operate as part of a photonic circuit, good transmission properties are required. This is ensured in the design process by adding a transmission constraint to the inverse design problem formulation, where the transmittance from the input to the output port is required to remain above $T_{\text{target}} = 0.9$, i.e. $T_{\text{Out}} > T_{\text{target}}$ implemented as described in [1]. It is noted that we find a tradeoff between the attainable quality factor for the device under design and the choice of $T_{\text{target}}$.

The figure of merit to be maximized in the design process is different for the two waveguides studied in this work. For the bowtie waveguide, extreme localization of the electric field at a single point in space, i.e. a small effective mode volume, is targeted by maximizing the electric field magnitude at a single point $r_0$ at the center of the design domain, $\Phi_1 = |E(r_0)|$. The reference device is designed to maximize the integral of the electric field intensity squared in the device material, $\Phi_2 = \int I_s(r) |E(r)|^4 d\Omega_D$, where $I_s$ is an indicator function that takes the value 0 in air and 1 in silicon. This figure of merit is proportional to the inverse of the TPA mode volume. As is standard in topology optimization, the design problem is recast as a mathematical optimization problem, which in turn is solved using the method of moving asymptotes [6]. The parameters used are listed in Table S1.

**Table S1**. Geometry, physics, and optimization parameters used in the modeling and topology optimization of the devices studied in this work.

| Geometry Parameters | | | | | | | | |
|---|---|---|---|---|---|---|---|---|
| $w_{\Omega_I}$ | $h_{\Omega_I}$ | $l_{\Omega_I}$ | $w_{\Omega_D}$ | $h_{\Omega_D}$ | $l_{\Omega_D}$ | $w_{wg}$ | $h_{wg}$ | $l_{wg}$ |
| 1600nm | 1600nm | 6200nm | 400nm | 110nm | 3100nm | 200nm | 110nm | 1600nm |
| **Physics and Optimization Parameters** | | | | | | | | |
| $\lambda$ | $c$ | $n_{air}$ | $n_{Si}$ | $r_f$ | $\beta$ | $\eta$ | $n_{iter}$ | |
| 1550nm | 3e8 m/s | 1.0 | 3.48 | 40nm | [4,8,16,32,64] | 0.5 | 200 | |

The optimized voxel-based geometries for both the reference and bowtie cavities are extracted and curve-fitted before being evaluated using a high-resolution mesh with second-order Nedelec elements to validate the optimized device performance.

## S2. Simulation of mode volumes and carrier diffusion

The calculated mode profiles for the nanobeam bowtie cavity, reference cavity, and conventional nanobeam cavity, are shown in Figure S2. The mode profiles of the reference cavity and the conventional nanobeam cavity [7] are very similar, but differ significantly from the mode profile of the nanobeam bowtie cavity, which shows a single hotspot with characteristic transverse dimension on the order of 10-20 nm.

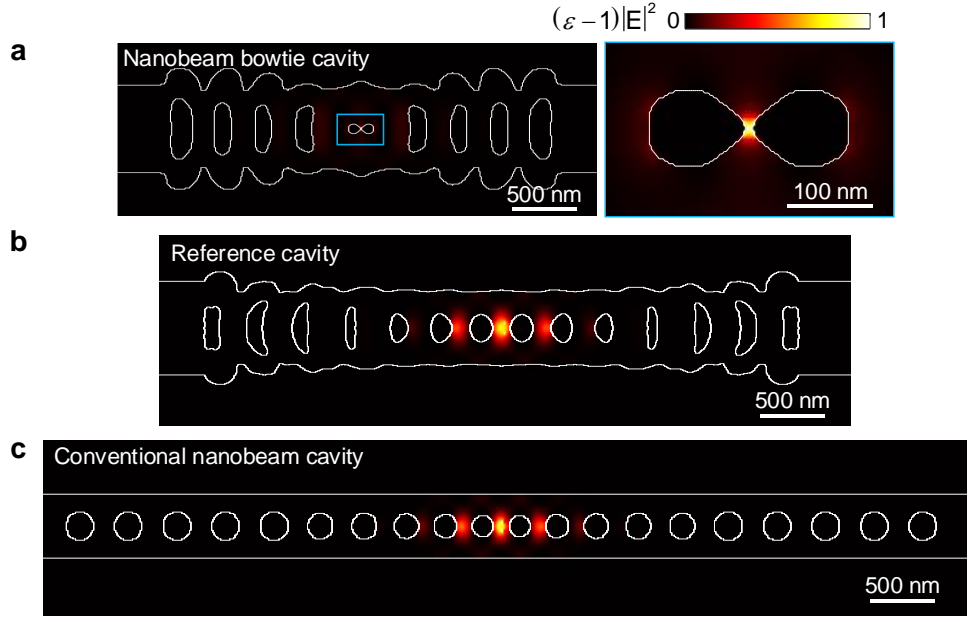

**Fig. S2**. **a**, **b**, **c**, Calculated fundamental modes of nanobeam bowtie cavity (**a**), reference cavity (**b**), and conventional nanobeam cavity (**c**) at the central plane.

As shown in Ref. [8], the dynamics of a microcavity depend on various characteristic mode volumes that reflect the nature of the light-matter interaction. In particular, we consider the following effective (linear) mode volume, $V_{\text{eff}}$, and the nonlinear mode volumes, including free-carrier absorption (FCA) mode volume, $V_{\text{FCA}}$, and two-photon absorption (TPA) mode volume, $V_{\text{TPA}}$, as expressed by

$$V_{\text{eff}} = \frac{\int_V n^2(\mathbf{r})|\mathbf{E}(\mathbf{r})|^2 \, dV}{n^2(\mathbf{r}_0)|\mathbf{E}(\mathbf{r}_0)|^2}, \tag{s5}$$

$$V_{\text{FCA}}^2 = \frac{\left(\int_V n^2(\mathbf{r})|\mathbf{E}(\mathbf{r})|^2 \, dV\right)^3}{\int_{V_{\text{Si}}} n^6(\mathbf{r})|\mathbf{E}(\mathbf{r})|^6 \, dV}, \tag{s6}$$

$$V_{\text{TPA}} = \frac{\left(\int_V n^2(\mathbf{r})|\mathbf{E}(\mathbf{r})|^2 \, dV\right)^2}{\int_{V_{\text{Si}}} n^4(\mathbf{r})|\mathbf{E}(\mathbf{r})|^4 \, dV}, \tag{s7}$$

where $n(\mathbf{r})$ and $\mathbf{E}(\mathbf{r})$ are the refractive index and electric field at position $\mathbf{r}$, respectively.

The mode volume, $V_{\text{eff}}$, is the standard linear mode volume entering into Purcell's formula for the emission rate of a dipole emitter placed at the position $\mathbf{r}_0$, which we here take to be at the center of the bowtie cavity [9]. This "linear" mode volume is rigorously given by the theory of quasinormal modes [10]. The nonlinear mode volume, $V_{TPA}$, governs the rate of TPA induced by the cavity mode. Notice that the integral in the denominator only extends over the solid material (Si); see Ref. [8] for details. The nonlinear mode volume, $V_{FCA}$, governs the carrier generation rate in the cavity mode, as well as the rate of free-carrier absorption in the cavity mode due to the carriers generated by TPA.

As shown in Table S2, the differences between the reference cavity and conventional nanobeam cavity in terms of effective mode volume, FCA mode volume, and TPA mode volume are below 10%. Both the effective mode volume and the FCA mode volume of the reference cavity and conventional nanobeam cavity are significantly larger than that of the bowtie cavity, while the TPA mode volume of these three types of cavities is comparable.

**Table S2.** Comparison of effective and nonlinear mode volumes of the nanobeam bowtie cavity (with a bowtie width of 12 nm), the reference cavity, and a conventional nanobeam cavity.

|  | $V_{eff}/(\lambda/2n)^3$ | $V_{FCA}/(\lambda/2n)^3$ | $V_{TPA}/(\lambda/2n)^3$ |
|---|---|---|---|
| Bowtie cavity | 0.39 | 4.2 | 10.3 |
| Reference cavity | 2.8 | 8.7 | 12.1 |
| Conventional nanobeam cavity | 2.9 | 8.2 | 11.0 |

Finally, we simulated the carrier relaxation processes for the bowtie cavity, the reference cavity, and the conventional nanobeam cavity for different surface recombination rates ($S = 10^4$, $10^5$, and $10^6$ cm/s), as shown in Fig. S3. The reference cavity and the conventional nanobeam cavity show comparable carrier relaxation processes for the same surface recombination rate, but differ significantly from the carrier dynamics of the nanobeam bowtie cavity. We also find that the carrier relaxation process is dominated by carrier diffusion when the surface recombination rate is lower than $10^5$ cm/s for all three types of cavities.

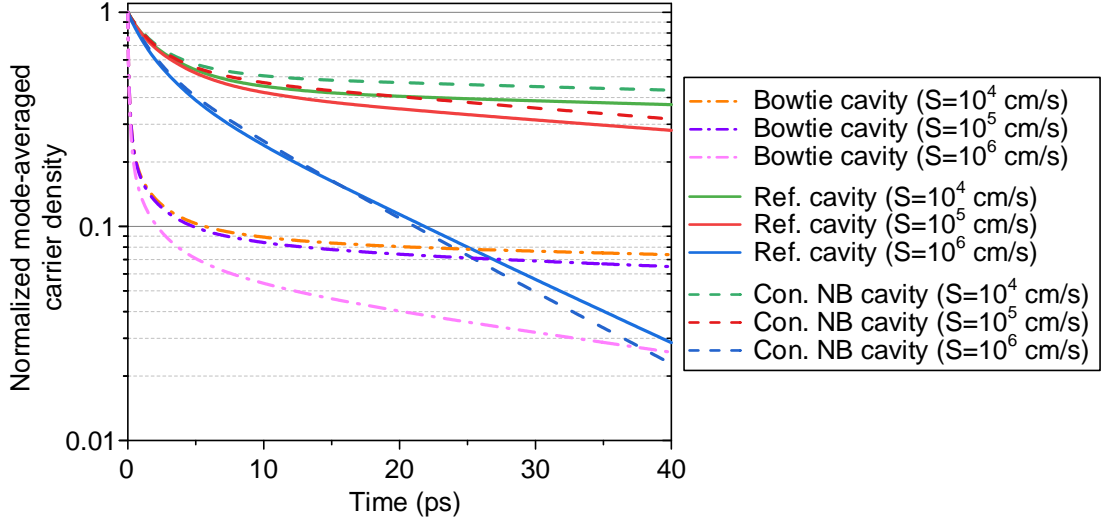

**Fig. S3.** Comparison of the carrier relaxation process in the nanobeam bowtie cavity (with a bowtie width of 12 nm), the reference cavity (Ref. cavity), and the conventional nanobeam cavity (Con. NB cavity) for different surface recombination rates ($10^4$, $10^5$, and $10^6$ cm/s).

## S3. Simulation of probe transmission dynamics

The probe transmission dynamics are calculated using temporal coupled-mode theory accounting for coherent effects, as described in Refs. [11] and [12]. The coherent interactions between the pump and probe fields leads to ultrafast parametric processes. Since the probe pulse energy is much smaller than the pump pulse energy, we treat the probe field as a small perturbation to the total field in the cavity. Thus, we apply a perturbation expansion to the injected power, the field inside the cavity, and the carrier density. The zeroth-order temporal dynamics, which describes the transmission dynamics of the pump field, can then be expressed through a set of ordinary differential equations [13]

$$\frac{d\tilde{a}_p}{dt} = -\left[i\delta_p + \kappa_{fc}N_1^{(0)} + \kappa_{tp}|a_p|^2 + \gamma\right]\tilde{a}_p + \sqrt{\gamma_c}\,s_p(t), \tag{s8-a}$$

$$\frac{dN_1^{(0)}}{dt} = -\frac{N_1^{(0)} - N_2^{(0)}}{\tau_{diff}} - \frac{N_1^{(0)}}{\tau_{slow}} + \frac{\beta_{TPA}c^2}{2\hbar\omega_p n^2 V_{FCA}^2}\left|\tilde{a}_p\right|^4, \tag{s8-b}$$

$$\frac{dN_2^{(0)}}{dt} = \frac{N_1^{(0)} - N_2^{(0)}}{\tau_{diff}}R_{12} - \frac{N_2^{(0)}}{\tau_{slow}}, \tag{s8-c}$$

Here, $\tilde{a}_p$ is the zeroth-order, complex, slowly varying field amplitude, $a_p(t) = \tilde{a}_p(t)e^{-i\omega_p t}$, normalized such that $|\tilde{a}|^2$ is the energy in the cavity, and $N_1$ and $N_2$ denote the carrier density in the cavity mode region and the background, respectively. The coefficient $R_{12}$ is the ratio of the effective volumes occupied by the different carrier densities $N_1$ and $N_2$, see Ref. [12] for details. The carrier decay in the region defined by the cavity mode proceeds through a fast diffusion process, with time constant $\tau_{diff}$, and a slower process, due to slow diffusion and non-radiative recombination, with time constant $\tau_{slow}$. The real and imaginary parts of $\kappa_{fc}$ represent FCA, $K_{FCA} = c\sigma/(2n)$, and free-carrier dispersion (FCD) and band filling, $K_{Car}$, respectively, with $c$ being the light speed in vacuum, $\sigma$ the absorption cross-section, and $n$ the material refractive index [13]. The real and imaginary parts of $\kappa_{tp}$ represent TPA, $K_{TPA} = \beta_{TPA}c^2/(2n^2 V_{TPA})$, and Kerr effect, $K_{Kerr} = \omega_0 c n_{2,Si}/(n^2 V_{TPA})$, respectively, with $\beta_{TPA}$ being the TPA coefficient, $\omega_0$ the resonant frequency of the cold cavity, and $n_{2,Si}$ the Kerr coefficient [13]. For a comprehensive discussion on carrier dynamics and diffusion, we refer to Refs. [11,12,13].

The rate equations for the first-order perturbations of the probe pulse, the idler pulse, and the two local carrier densities are given by

$$\frac{d\tilde{a}_s}{dt} = -\left[i\delta_s \tilde{a}_s + \kappa_{fc}\left(N_1^{(0)}\tilde{a}_s + n_1^*\tilde{a}_p\right) + \kappa_{tp}\left(2|\tilde{a}_p|^2 \tilde{a}_s + \tilde{a}_p^2 \tilde{a}_i^*\right) + \gamma \tilde{a}_s\right] + \sqrt{\gamma_c}\,s_s(t) \quad \text{(s9-a)}$$

$$\frac{d\tilde{a}_i}{dt} = -\left[i\delta_i \tilde{a}_i + \kappa_{fc}\left(N_1^{(0)}\tilde{a}_i + n_1 \tilde{a}_p\right) + \kappa_{tp}\left(2|\tilde{a}_p|^2 \tilde{a}_i + \tilde{a}_p^2 \tilde{a}_s^*\right) + \gamma \tilde{a}_i\right] \quad \text{(s9-b)}$$

$$\frac{dn_1}{dt} = i\delta_{ps} n_1 - \frac{n_1 - n_2}{\tau_{diff}} - \frac{n_1}{\tau_{slow}} + \frac{\beta_{TPA}c^2}{\hbar \omega_p n^2 V_{FCA}^2}\left[\tilde{a}_p^* \tilde{a}_i + \tilde{a}_p \tilde{a}_s^*\right]|\tilde{a}_p|^2 \quad \text{(s9-c)}$$

$$\frac{dn_2}{dt} = i\delta_{ps} n_2 + \frac{n_1 - n_2}{\tau_{diff}}R_{12} - \frac{n_2}{\tau_{slow}} \quad \text{(s9-d)}$$

where $\delta_x = \omega_0 - \omega_x$, with indices $i$, $p$, and $s$ denoting idler, pump, and probe, respectively. Also, $\delta_{ps} = \omega_p - \omega_s$ is the beat frequency between the pump and probe at which the slowly varying first-order carrier density amplitudes $n_1$ and $n_2$ oscillate. Here, $s_p(t)$ and $s_s(t)$ are the amplitudes of the incoming pump and probe waves, respectively. The coefficients $\gamma$ and $\gamma_c$ are related to the total cavity Q factor as $\gamma = \omega_0/2Q_t$ and the cavity coupling Q factor as $\gamma_c = \omega_0/2Q_c$, respectively. The final output energy flux is $f_x(t) = \gamma_c |\tilde{a}_x(t)|^2$, $x \in \{i, p, s\}$.

**Table S3.** Key parameters used for simulations of nanobeam bowtie cavity and reference cavity.

| Parameters | Nanobeam bowtie cavity | Reference cavity |
|---|---|---|
| $V_{FCA}$ | $0.033 \times 10^{-18}$ m³ | $0.097 \times 10^{-18}$ m³ |
| $V_{TPA}$ | $0.113 \times 10^{-18}$ m³ | $0.138 \times 10^{-18}$ m³ |
| $\tau_{diff}$ | 0.6 ps | 6 ps |
| $\tau_{slow}$ | 100 ps | 100 ps |
| $R_{12}$ | 1/10 | 1/1.3 |
| $Q_t$ | 700 | 1200 |
| $\Delta\tau_{pump}$ | 0.93 ps | 1.12 ps |
| | Values below are identical for the bowtie and the reference cavity | |
| $\beta_{TPA}$ | $9.95 \times 10^{-12}$ m/W | |
| $n_2$ | $4 \times 10^{-18}$ m²/W | |
| $n$ | 3.45 | |
| $K_{Car}$ | $0.89 \times 10^{-12}$ m³/s | |
| $\sigma$ | $1.45 \times 10^{-21}$ m² | |
| $\Delta\tau_{probe}$ | 5.0 ps | |
| $E_{pump}$ | 255 fJ/pulse | |
| $E_{probe}$ | 0.64 fJ/pulse | |

Figure S4 shows the calculated probe transmission dynamics of the nanobeam bowtie cavity and reference cavity with and without the parametric process, i.e., the four-wave mixing (FWM) effect, by using the parameters given in Table S3. When the FWM effect is included, the calculated results (Fig. S4 a(i), b(i), and c(i)) agree well with the corresponding measured results (Fig. 4 a(i), b(i), and c(i) in the main text). Without the FWM effect, the amplitude of the fast probe transmission dynamics is reduced significantly for both bowtie and reference cavity, while the slow recovery dynamics are nearly unaffected. This indicates that the fast probe transmission oscillations are dominated by the FWM effect, while the slow recovery tails depend on the cavity resonance shift. The red dashed curves show the cavity resonance shift. We see that the FWM effect only slightly affects the cavity resonance shift for both the bowtie and reference cavities.

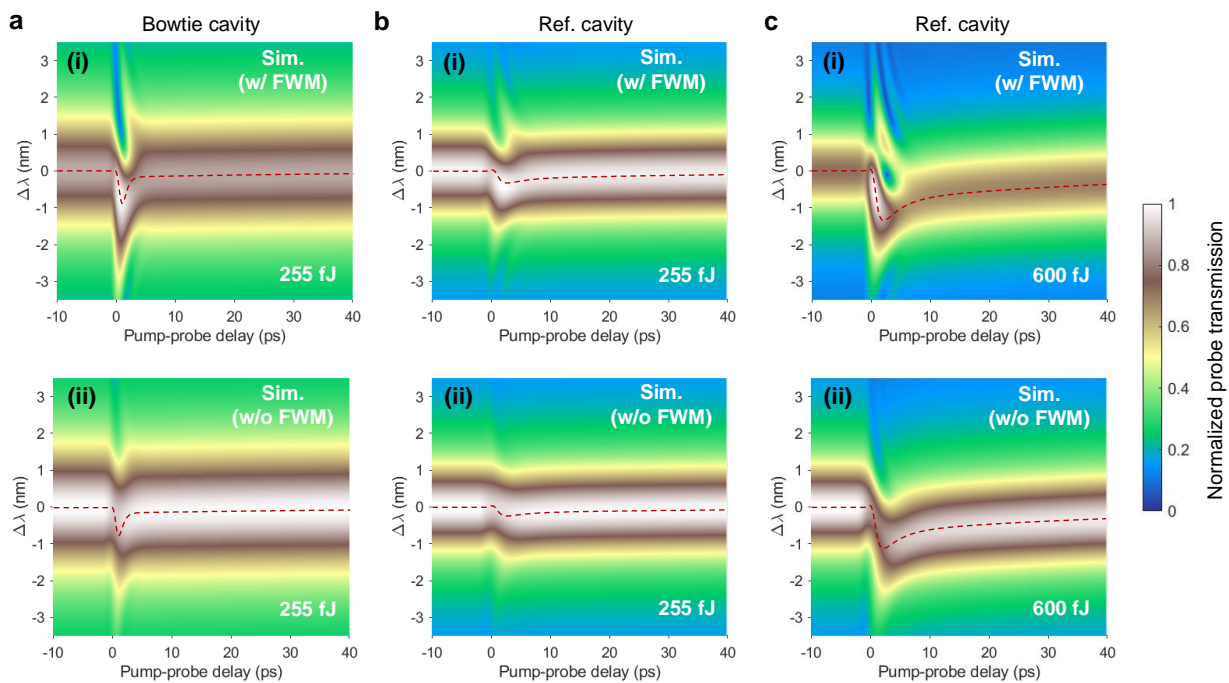

**Fig. S4. a**, Calculated probe transmission dynamics of the nanobeam bowtie cavity with (i) and without (ii) FWM at a pump pulse energy of 255 fJ. **b**, **c**, Calculated probe transmission dynamics of the reference cavity with (i) and without (ii) FWM at a pump pulse energy of 255 fJ (**b**) and 600 fJ (**c**), respectively. The red dashed curves represent the corresponding cavity resonance shifts.

## S4. Experimental setup and heterodyne pump-probe technique

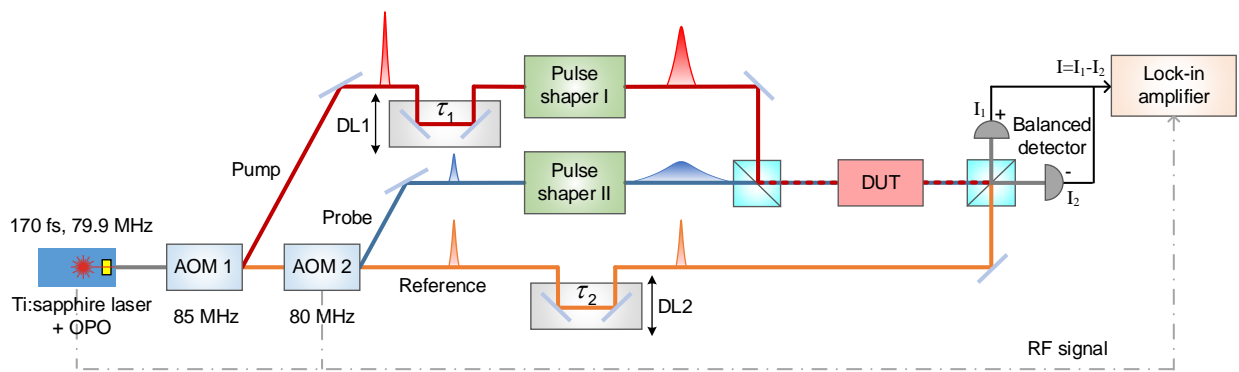

**Fig. S5.** Sketch of heterodyne pump-probe experimental setup. OPO: optical parametric oscillator; AOM: acousto-optical modulators; DL: delay lines; RF signal: radio-frequency signal; DUT: device under test.

Figure S5 shows the heterodyne pump-probe experimental setup, based on the principles described in Ref. [14]. A Ti:Sapphire pulsed laser (Spectra-Physics Maitai) in combination with an optical parametric oscillator (Spectra-Physics Inspire HP) generates ~170-fs pulses centered around 1.54 μm with a repetition rate of 79.9 MHz ($\Omega_0/2\pi$). Two acousto-optical modulators were used to deflect part of the reference beam into pump and probe beams with frequency shifts of 85 MHz ($\Omega_1/2\pi$) and 80 MHz ($\Omega_2/2\pi$) relative to the reference pulse, respectively. A fiber-coupled waveshaper (Finisar Waveshaper 4000S/X) was used to reshape the spectra of the pump and probe pulses through Gaussian filters and compensate for fiber-induced dispersion. As the bandwidth of the applied Gaussian filters was much narrower than the bandwidth of the laser spectrum, the spectrum was nearly constant across the passband of the filters, thus carving out a Gaussian spectrum. Therefore, the pump and probe pulses are close to transform-limited during the measurements.

Two adjustable delay lines were inserted into the pump and reference paths to adjust the relative arrival time between the pump, probe, and reference pulses. The pump and probe pulses were TE-polarized. The combined pump and probe pulses were coupled into the waveguide through a grating coupler with a coupling loss of −6 dB per facet. The probe signal was selected using a heterodyne cross-correlation technique by combining it with the reference pulse via a 50:50 beam splitter and using a lock-in amplifier (Stanford Research 844) locked to the beating frequency ($\left(\Omega_2 - \Omega_0\right)/2\pi$) between the laser repetition rate and the modulation frequency of AOM 2. The idler signal, instead, was extracted with a locked frequency of $\left(2\Omega_1 - \Omega_2 - \Omega_0\right)/2\pi = 10.1$ MHz, as shown in Figure S6.

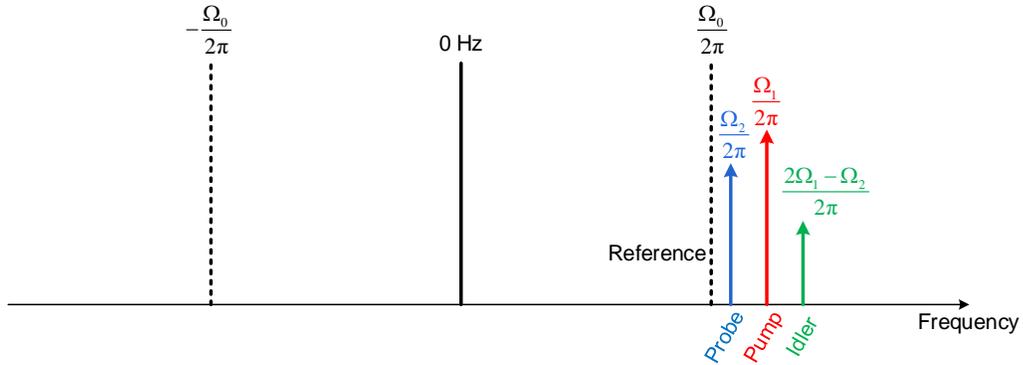

**Fig. S6.** Radio frequency distribution of the heterodyne signals. Here we only show the 0$^\text{th}$-order pump and probe signals, as well as the generated idler by four-wave mixing between the 0$^\text{th}$-order pump and probe.

In the balanced heterodyne detection system, the photocurrents generated at the two ports of the balanced photodetector are proportional to

$$I_1(t) =$$

$$\int_{-T_{res}/2}^{T_{res}/2} \begin{bmatrix} E_p(t-\tau_1)\exp(-j\Omega_1(t-\tau_1)) + E_s(t)\exp(-j\Omega_2 t) + \\ E_i(t-\tau_1)\exp(-j(2\Omega_1-\Omega_2)(t-\tau_1)) + E_r(t-\tau_2)\exp(-j\Omega_0(t-\tau_2)) \end{bmatrix}$$

$$\times \begin{bmatrix} E_p^*(t-\tau_1)\exp(j\Omega_1(t-\tau_1)) + E_s^*(t)\exp(j\Omega_2 t) + \\ E_i^*(t-\tau_1)\exp(j(2\Omega_1-\Omega_2)(t-\tau_1)) + E_r^*(t-\tau_2)\exp(j\Omega_0(t-\tau_2)) \end{bmatrix} dt \quad \text{(s10)}$$

and

$$I_2(t) =$$

$$\int_{-T_{res}/2}^{T_{res}/2} \begin{bmatrix} E_p(t-\tau_1)\exp(-j\Omega_1(t-\tau_1)) + E_s(t)\exp(-j\Omega_2 t) + \\ E_i(t-\tau_1)\exp(-j(2\Omega_1-\Omega_2)(t-\tau_1)) - E_r(t-\tau_2)\exp(-j\Omega_0(t-\tau_2)) \end{bmatrix},$$

$$\times \begin{bmatrix} E_p^*(t-\tau_1)\exp(j\Omega_1(t-\tau_1)) + E_s^*(t)\exp(j\Omega_2 t) + \\ E_i^*(t-\tau_1)\exp(j(2\Omega_1-\Omega_2)(t-\tau_1)) - E_r^*(t-\tau_2)\exp(j\Omega_0(t-\tau_2)) \end{bmatrix} dt \quad \text{(s11)}$$

where $E_p(t-\tau_1)\exp(-j\Omega_1(t-\tau_1))$, $E_s(t)\exp(-j\Omega_2 t)$, $E_i(t-\tau_1)\exp(-j(2\Omega_1-\Omega_2)(t-\tau_1))$, and $E_r(t-\tau_2)\exp(-j\Omega_0(t-\tau_2))$ are the pump, probe, idler, and reference signals, respectively. $T_{res}$ is the time resolution of the photodetector which fulfills the relation $T_{Pulse\ width} \ll T_{res} \ll 2\pi/(\Omega_2-\Omega_0)$. The output photocurrent of the balanced detector is the difference of its two branches, i.e. $I(t) = I_1(t) - I_2(t)$. The signal, $I(t)$, is then injected into the lock-in amplifier. Thus, multiplying with the lock-in signal (extracting the probe signal), $A_0\exp(-j(\Omega_2-\Omega_0)t)$, with $A_0$ being a constant, and selecting only the DC component (the AC component has been removed by the low-pass filter in the amplifier), the finally detected signal is proportional to

$$P_{Lock-in}(\tau_2) \propto 2A_0 \int_{-T_{res}/2}^{T_{res}/2} |E_r(t-\tau_2)E_s^*(t)| dt. \quad \text{(s12)}$$

When the reference pulse width is very narrow, i.e. much narrower than the probe pulse width, which is the case in this work, the probe pulse can be viewed as an impulse function, thereby the detected signal is proportional to the amplitude of the probe field, $|E_s^*(\tau_2)|$, at a probe-reference delay of $\tau_2$.

## S5. Transmission spectra

Figure S7 shows the transmission spectra of the nanobeam bowtie cavity (Fig. S7a) and the reference cavity (Fig. S7b), excluding the coupling loss of the grating coupler. The total quality factor of the bowtie (reference) cavity is 700 (1200) according to a Lorentzian fit, and the peak transmission is as high as $-0.82$ dB ($-0.47$ dB), indicating an intrinsic quality factor of 7800 (23000). Both the quality factors and peak transmissions of the bowtie and reference cavity show good agreement with simulations. The ripples in the measured spectra originate from the reflections between the grating coupler and the cavity.

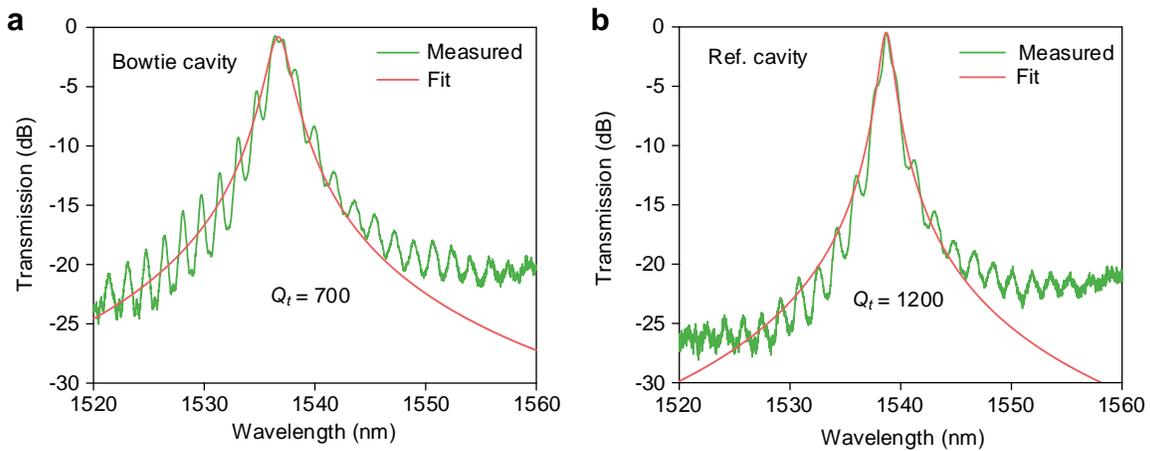

**Fig. S7. a – b**, Transmission spectra of the nanobeam bowtie cavity (**a**) and the reference cavity (**b**).

## S6. Idler signal of bowtie cavity

To further confirm that the ultrafast dynamics are caused by four-wave mixing, we extracted the idler signal from the output mixed signal, as shown in Figure S8a. The idler light only exists within a very short pump-probe delay range, and the intensity of the idler light decreases with the increase of the probe detuning, in good agreement with our theoretical results (Fig. S8b). The temporal extent of the idler signal depends on the lifetime of the intracavity field, that is, the larger value of the photon lifetime and the pump pulse width. The intensity of the idler light depends on the intensity of the intracavity pump field and probe field, as well as the phase matching conditions between the pump, the probe, and the oscillating free carriers.

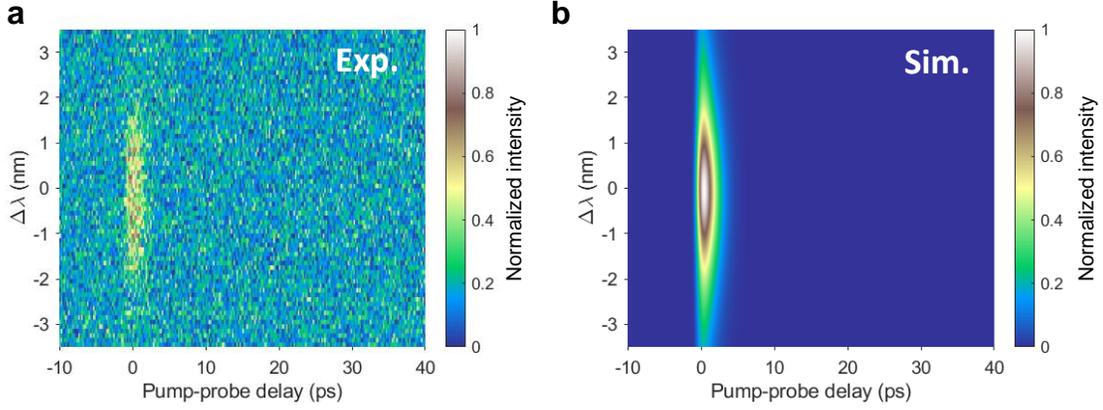

**Fig. S8. a – b**, Measured (**a**) and calculated (**b**) idler signals as a function of the pump-probe delay and probe wavelength. The pump wavelength coincides with the peak resonance of the cold cavity. The pump pulse energy is 255 fJ at the input waveguide.

## S7. Impact of nonlinear absorption loss

To study the impact of nonlinear absorption losses on the probe transmission dynamics of the nanobeam bowtie cavity, we calculated the transmission dynamics when considering different nonlinear effects, as shown in Figure S9. Here, we define a figure of merit, FOM = $\Delta P_X/\Delta P_0$, as the influence of $X$ (TPA, FCA, Kerr) effect on the extinction ratio. We see that during the parametric suppression (amplification) process, the influence of TPA and FCA on the extinction ratio are 3.5% (10%) and 6.1% (13.5%), respectively, meaning that the impact of the nonlinear losses on the parametric process is relatively small. In addition, the influence of these nonlinear absorption losses on the slow-varying tail is almost negligible. The Kerr effect only slightly changes the extinction ratio.

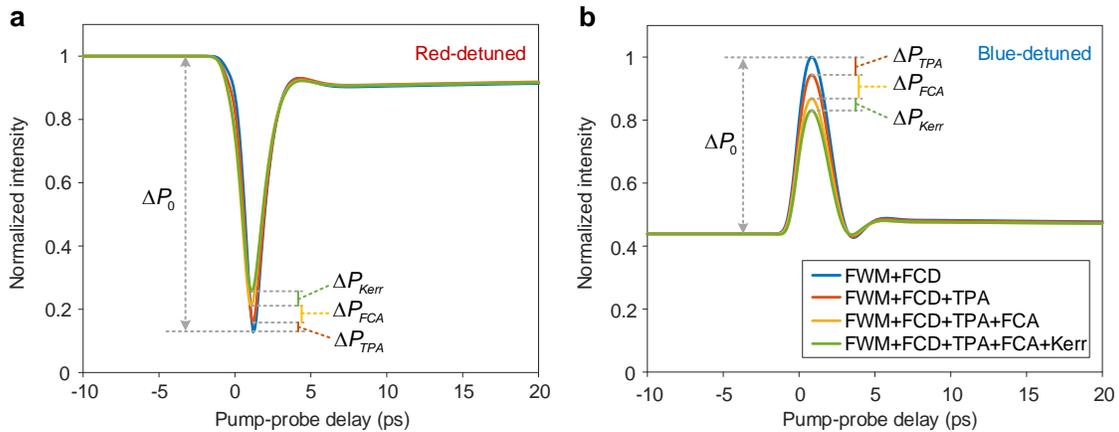

**Fig. S9. a – b**, Impact of nonlinear absorption losses on the transmission dynamics during parametric suppression (**a**) and amplification (**b**) of the nanobeam bowtie cavity. The pump pulse energy is 255 fJ at the input waveguide.

## S8. Impact of pump pulse width

Figure S10 shows the probe transmission dynamics for different pump pulse widths for both red-detuned (Fig. S10a) and blue-detuned (Fig. S10b) probes. The results illustrate that for a given pump pulse energy, the extinction ratio has an optimum for a specific width of the pump pulse. This optimized pump pulse width is determined by the photon lifetime of the cavity. When the pump pulse is shorter than the cavity photon lifetime, only part of the spectrum of the pump pulse overlaps with the cavity spectrum, and the coupling efficiency is reduced, thereby weakening the nonlinearities in the cavity. On the contrary, when the pump pulse is longer than the photon lifetime, although the dwell time of pump light in the cavity increases, the peak intensity of the intracavity field decreases, resulting in a decrease in the extinction ratio but an increase in the temporal extent of the switching window.

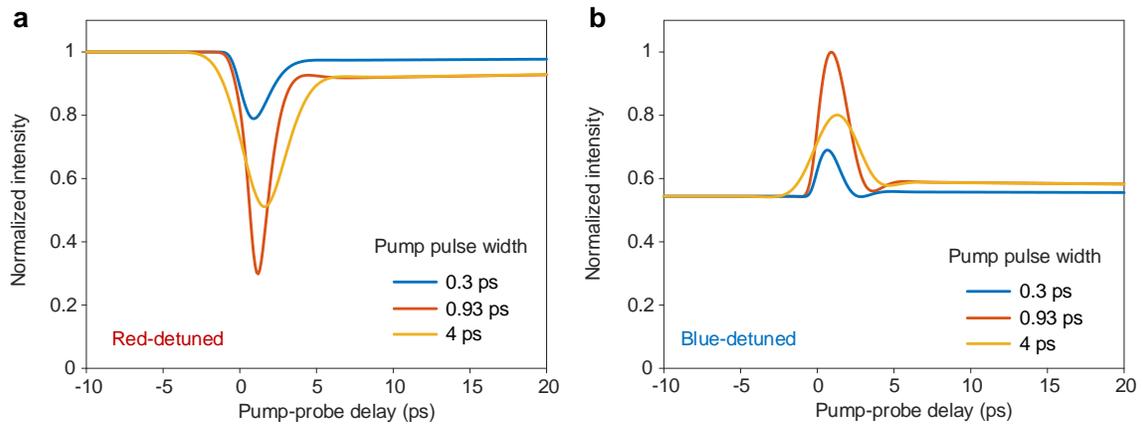

**Fig. S10**. **a** – **b**, Calculated probe transmission dynamics for a red-detuned (a) and blue-detuned (b) probe for different pump pulse widths (0.3 ps, 0.93 ps, and 4 ps). The pump pulse energy is fixed at 255 fJ, and the probe pulse width is fixed at 5 ps.

## References

1. Christiansen, R. E., "Inverse design of optical mode converters by topology optimization: tutorial," J. Opt. **25**, 083501 (2023).

2. Jin, J.-M., *The finite element method in electromagnetics* (John Wiley & Sons, 2015).

3. Christiansen, R. E. & Sigmund, O., "Compact 200 line MATLAB code for inverse design in photonics by topology optimization: tutorial," J. Opt. Soc. Am. B **38**, 510 (2021).

4. Wang, F., Lazarov, B. S. & Sigmund, O., "On projection methods, convergence and robust formulations in topology optimization," Struct. Multidisc. Optim. **43**, 767-784 (2010).

5. Zhou, M., Lazarov, B. S., Wang, F. & Sigmund, O., "Minimum length scale in topology optimization by geometric constraints," Comput. Methods Appl. Mech. Eng. **293**, 266-282 (2015).

6. Svanberg, K., "A class of globally convergent optimization methods based on conservative convex separable approximations," SIAM J. Optim. **12**, 555-573 (2002).

7. Deotare, P. B., et al., "High quality factor photonic crystal nanobeam cavities," Appl. Phys. Lett. **94**, 121106 (2009).

8. Barclay, P. E., Srinivasan, K. & Painter, O., "Nonlinear response of silicon photonic crystal microresonators excited via an integrated waveguide and fiber taper," Opt. Express **13**, 801-820 (2005).

9. Albrechtsen, M., Vosoughi Lahijani, B. & Stobbe, S., "Two regimes of confinement in photonic nanocavities: bulk confinement versus lightning rods," Opt. Express **30**, 15458-15469 (2022).

10. Kristensen, P. T., Van Vlack, C. & Hughes, S., "Generalized effective mode volume for leaky optical cavities," Opt. Lett. **37**, 1649-1651 (2012).

11. Colman, P., Lunnemann, P., Yu, Y. & Mork, J., "Ultrafast coherent dynamics of a photonic crystal all-optical switch," Phys. Rev. Lett. **117**, 233901 (2016).

12. Lunnemann, P., Yu, Y., Joanesarson, K. & Mørk, J., "Ultrafast parametric process in a photonic-crystal nanocavity switch," Phys. Rev. A **99**, 053835 (2019).

13. Yu, Y., et al., "Switching characteristics of an InP photonic crystal nanocavity: Experiment and theory," Opt. Express **21**, 31047-31061 (2013).

14. Mecozzi, A. & Mørk, J., "Theory of heterodyne pump–probe experiments with femtosecond pulses," J. Opt. Soc. Am. B **13**, 2437-2452 (1996).